\documentclass[conference,10pt,letterpaper]{IEEEtran}

\ifCLASSINFOpdf
   \usepackage[pdftex]{graphicx}
\else
\fi

\usepackage{comment}

\usepackage{amsmath}
\usepackage{mathtools}

\usepackage{acronym}

\usepackage{amsfonts}
\usepackage{amsthm}
\usepackage{thmtools}

\usepackage{siunitx}
\usepackage{booktabs}

\usepackage{mleftright}
\usepackage{bm}
\usepackage{bbm}

\usepackage{tikz}
\usepackage{pgfplots}

\usepackage[textsize=tiny]{todonotes} %

\usepackage[font=footnotesize, skip=8pt]{caption}
\ifCLASSOPTIONcompsoc
  \usepackage[caption=false,font=normalsize,labelfont=sf,textfont=sf]{subfig}
\else
  \usepackage[caption=false,font=footnotesize]{subfig}
\fi

\newcommand\blfootnote[1]{%
  \begingroup
  \renewcommand\thefootnote{}\footnote{#1}%
  \addtocounter{footnote}{-1}%
  \endgroup
}

\usepackage{graphicx}
\usepackage{color}
\usepackage{pgf, tikz, pgfplots}
\usetikzlibrary{shapes,arrows.meta, calc, matrix, patterns}
\usepgfplotslibrary{groupplots}

\usepackage[ruled,vlined, linesnumbered]{algorithm2e}
\let\oldnl\nl%
\newcommand{\nonl}{\renewcommand{\nl}{\let\nl\oldnl}}%

\SetCommentSty{mycommfont}

\definecolor{KITgreen}{rgb}{0,.59,.51}
\definecolor{KITpalegreen}{RGB}{130,190,60} 
\definecolor{KITblack}{rgb}{0,0,0}
\definecolor{KITblue}{rgb}{.27,.39,.66}
\definecolor{KITred}{rgb}{.63,.13,.13}
\definecolor{KITpurple}{rgb}{.64,.06,.48}
\definecolor{KITcyan}{rgb}{.14,.63,.87}
\definecolor{KITyellow}{rgb}{.98,.89,0}
\definecolor{KITorange}{rgb}{.87,.60,.10}

\definecolor{EMBPcolor}{rgb}{0,.59,.51}
\definecolor{BPcolor}{rgb}{.64,.06,.48}
\definecolor{EMNBPcolor}{RGB}{130,190,60} 
\definecolor{EMBPLCcolor}{rgb}{.63,.13,.13}
\definecolor{EMBPDiraccolor}{rgb}{.14,.63,.87}
\definecolor{VAEcolor}{rgb}{.87,.60,.10}
\colorlet{RNNcolor}{KITblue}
\colorlet{ViterbiNetcolor}{KITcyan}

    \acrodef{APP}[APP]{a posteriori probability}
    \acrodef{AWGN}[AWGN]{additive white Gaussian noise}
    \acrodef{BP}[BP]{belief propagation}
    \acrodef{BPSK}[BPSK]{binary phase-shift keying}
    \acrodef{BER}[BER]{bit error rate}
    \acrodef{BMD}[BMD]{bit-metric decoder}
    \acrodef{BMI}[BMI]{bitwise mutual information}
    \acrodef{BICM}[BICM]{bit-interleaved coded modulation}
    \acrodef{CMA}[CMA]{constant modulus algorithm}
    \acrodef{CSI}[CSI]{channel state information}
    \acrodef{DNN}[DNN]{deep neural network}
    \acrodef{ELBO}[ELBO]{evidence lower bound}
    \acrodef{EM}[EM]{expectation maximization}
    \acrodef{FEC}[FEC]{forward error correction}
    \acrodef{FIR}[FIR]{finite impulse response}
    \acrodef{iid}[i.i.d.]{independent and identically distributed}
    \acrodef{ISI}[ISI]{inter-symbol interference}
    \acrodef{KL}[KL]{Kullback-Leibler}
    \acrodef{LLR}[LLR]{log-likelihood ratio}
    \acrodef{MAP}[MAP]{maximum a posteriori}
    \acrodef{ML}[ML]{maximum likelihood}
    \acrodef{MMSE}[LMMSE]{linear minimum mean squared error}
    \acrodef{MSE}[MSE]{mean squared error}
    \acrodef{NBP}[NBP]{neural belief propagation}
    \acrodef{NN}[NN]{neural network}
    \acrodef{pdf}[PDF]{probability density function}
    \acrodef{pmf}[PMF]{probability mass function}
    \acrodef{RNN}[RNN]{recurrent neural network}
    \acrodef{SNR}[SNR]{signal-to-noise ratio}
    \acrodef{SPA}[SPA]{sum-product algorithm}
    \acrodef{VAELE}[VAE-LE]{variational autoencoder-based linear equalizer}

\newcommand{\Estep}{\mbox{E-step}}
\newcommand{\Mstep}{\mbox{M-step}}

\pgfplotsset{compat=newest}
\begin{document}
\title{Optimization of Iterative Blind Detection based on Expectation Maximization and Belief Propagation}

\author{\IEEEauthorblockN{Luca Schmid\IEEEauthorrefmark{1},
Tomer Raviv\IEEEauthorrefmark{2}, Nir Shlezinger\IEEEauthorrefmark{2}, and
Laurent Schmalen\IEEEauthorrefmark{1}}
\IEEEauthorblockA{\IEEEauthorrefmark{1}Communications Engineering Lab, Karlsruhe Institute of Technology, Hertzstr. 16, 76187 Karlsruhe, Germany\\
\IEEEauthorrefmark{2}School of ECE, Ben-Gurion University of the Negev, Be'er-Sheva, Israel\\
Email: \IEEEauthorrefmark{1}\texttt{first.last@kit.edu},
\IEEEauthorrefmark{2}\texttt{tomerraviv95@gmail.com}, \texttt{nirshl@bgu.ac.il}}
}
\markboth{Submitted version, \today}%
{Schmid \MakeLowercase{\textit{et al.}}: XXX}

\maketitle
\begin{abstract}
We study  iterative blind symbol detection for block-fading linear inter-symbol interference channels. 
Based on the factor graph framework, we design a joint channel estimation and detection scheme that combines the expectation maximization (EM) algorithm and the ubiquitous belief propagation (BP) algorithm.
Interweaving the iterations of both schemes significantly reduces the EM algorithm's computational burden while retaining its excellent performance.
To this end, we apply simple yet effective model-based learning methods to find a suitable parameter update schedule by introducing momentum in both the EM parameter updates as well as in the BP message passing.
Numerical simulations verify that the proposed method can learn efficient schedules that generalize well and even outperform coherent BP detection in high signal-to-noise scenarios. 
\blfootnote{This work has received funding in part from the European Research Council (ERC) under the European Union’s Horizon 2020 research and innovation programme (grant agreement No. 101001899), in part from the German
Federal Ministry of Education and Research (BMBF) within the project Open6GHub (grant agreement 16KISK010), and in part from the Israeli Ministry of Science and Technology.}%
\end{abstract}
\begin{IEEEkeywords}
    \noindent Factor graphs, joint detection, model-based learning, expectation maximization, belief propagation, 6G.
\end{IEEEkeywords}

\IEEEpeerreviewmaketitle

\section{Introduction}
We study the application of iterative receivers for joint channel estimation and symbol detection on block-fading linear \ac{ISI} channels.
Most practical communication systems estimate the current channel state based on a priori known pilot symbols. However, in scenarios with short block length transmission, like low-latency communications in 5G/6G or Internet of Things systems, the pilots occupy a significant part of the short transmission blocks and thus significantly reduce the throughput.
This motivates considering \emph{blind} receivers that do not rely on pilot symbols, but estimate the channel state from the statistics of the received signal.

Optimum blind estimation and detection based on the \ac{ML} criterion~\cite{ghosh_maximum-likelihood_1992} is competitive in terms of detection performance with coherent detectors and fully available \ac{CSI} but suffers from high computational complexity. 
Alternatively, practical blind detectors with reduced complexity, like the \ac{CMA}~\cite{godard_self-recovering_1980} or variational methods~\cite{lauinger_blind_2022} are still consistently outperformed by coherent detection schemes.

Iterative receivers bear the potential of approximating \ac{ML}-based joint estimation and detection with an attractive performance-complexity tradeoff~\cite{worthen_unified_2001}. To this end, we study the framework of factor graphs and the \ac{BP} algorithm for joint estimation and detection. One major challenge is the handling of continuous latent variables in the factor graph which typically leads to intractable integrals in the \ac{BP} update equations.
Approaches in the literature combatting this issue vary from quantized \ac{BP} messages over particle filters~\cite{loeliger_remarks_2003} to parametrized canonical distributions~\cite{worthen_unified_2001}. For instance, \cite{liu_joint_2009} approximates the continuous probability distributions by mixtures of Gaussian distributions.
A promising approach was considered in~\cite{eckford_channel_2004, dauwels_expectation_2005}, where the \ac{EM} algorithm was used as a message passing algorithm in the factor graph to handle continuous variables.

Inspired by this idea, we integrate the \ac{EM} algorithm into our \ac{BP}-based detector for joint estimation and detection. In contrast to~\cite{dauwels_expectation_2005}, we do not apply \ac{EM} as a local message update rule but instead perform global \ac{EM} parameter updates which turn out to be more efficient in terms of complexity. 
The combination of two iterative schemes like the \ac{EM} and \ac{BP} algorithm raises the question of suitable update schedules which have a significant impact on both performance and complexity. 
To this end, we unfold the iterations of the \ac{EM} and \ac{BP} algorithm as a form of model-based deep learning~\cite{shlezinger_model-based_2023,shlezinger2022model} and introduce momentum in the updates by a convex combination with former values. 
Thereby, the schedule is fully defined by a set of continuous momentum weights which can be learned in an end-to-end manner via gradient-based optimization.
Our numerical experiments show that the optimized schedule can substantially reduce the number of necessary \ac{EM} steps by allowing parallel parameter updates while preserving the stability of the \ac{EM} algorithm. In a similar manner, momentum can improve the convergence behavior of \ac{BP} which leads to an improved detection performance.

\section{System Model and Preliminaries}\label{sec:channel_model_and_prelims}
\subsection{Channel Model}\label{subsec:system_model}
We study the transmission over a block-fading channel in the digital baseband which is impaired by linear \ac{ISI} and \ac{AWGN}~\cite{proakis_digital_2007}.
The channel state is constant for a block of $N$ consecutively transmitted information symbols ${\bm{c}\in\mathcal{M}^N}$, where each symbol $c_n$ is sampled independently and uniformly from a multilevel constellation ${\mathcal{M} = \{\text{c}_i \in \mathbb{C}, i=1,\ldots,M \}}$. 
The receiver observes the sequence
\begin{equation*}
    \bm{y} = \bm{H}\bm{c} + \bm{w}, \qquad \bm{H} \in \mathbb{C}^{N+L \times N}
    \label{eqn:Channel}
\end{equation*}
where $\bm{H}$ is the band-structured Toeplitz matrix that represents the convolution of $\bm{c}$ with the impulse response of the linear \ac{ISI} channel ${\bm{h} = (h_0,\ldots,h_L)^{\mathrm{T}} \in \mathbb{C}^{L+1}}$. The vector~$\bm{w}$ contains independent noise samples~${w_n \sim \mathcal{CN}(0,\sigma^2)}$ from a complex circular Gaussian distribution. 
For each transmission block, the channel is thus fully characterized by the parameter vector~$\bm{\theta} := (h_0, \ldots, h_L, \sigma^2 )^{\rm T}$ of length ${L+2}$, which is independently sampled in each block. %
We define the \ac{SNR} at the receiver as
\begin{equation*}
    \mathsf{snr} := \frac{\lVert \bm{h} \rVert^2 \cdot \mathbb{E}_{\bm{c}
    } \mleft\{ \lVert \bm{c}\rVert^2 \mright\}}{\mathbb{E}_{\bm{w}} \mleft\{ \lVert \bm{w} \rVert^2 \mright\}}.
\end{equation*}

We study \emph{blind} channel estimation and joint symbol detection, i.e., the parameters $\bm{\theta}$ are unknown to both the transmitter and the receiver. The goal is to infer the channel state~$\bm{\theta}$ and the transmitted symbols~$\bm{c}$ from the observation~$\bm{y}$.

\subsection{Factor Graph-based Symbol Detection}\label{sec:ufg}
As a basis for iterative reception, we consider the powerful framework of factor graphs and the \ac{BP} algorithm~\cite{worthen_unified_2001,kschischang_factor_2001}. We focus on a detection algorithm with particularly low computational complexity~\cite{colavolpe_siso_2011}, which we introduce in the following. For a more detailed elaboration of this detector, we refer the reader to~\cite{colavolpe_siso_2011,schmid_low-complexity_2022}.

\Ac{MAP} detection is based on the \ac{APP} distribution~${P(\bm{c}|\bm{y})}$. Assuming uniformly distributed transmit symbols, the latter can be written as
\begin{equation}
    P(\bm{c}|\bm{y}) \propto p(\bm{y|c}) \propto \exp\mleft( {2{\text{Re}}\mleft\{ \bm{c}^{\text H} \bm{x} \mright\} -\bm{c}^{\text H} \bm{G} \bm{c}\over \sigma^2}  \mright), \label{eq:expanded_likelihood}
\end{equation}
by defining ${ \bm{x} := \bm{H}^{\text H} \bm{y}}$ and ${\bm{G} := \bm{H}^{\text H} \bm{H}}$
as the matched filtered versions of the observation and the channel matrix, respectively.
The proportionality $\propto$ in~\eqref{eq:expanded_likelihood} indicates that the terms are only differing in a factor independent of $\bm{c}$.
Equation~\eqref{eq:expanded_likelihood} can be written in factorized form
\begin{equation}
    P(\bm{c} | \bm{y}) \propto \prod\limits_{n=1}^{N} \left [ {F_n(c_n) }
                \prod\limits_{m=1}^{n-1}  {I_{nm}(c_n,c_m) } \right ],
\label{eq:Ungerboeck_factorization}
\end{equation}
where we introduced the local factors
\begin{align}
    F_n(c_n) &:=  \exp \mleft \{ \frac{1}{\sigma^2} {\text{Re}} \mleft\{ 2 x_n c_n^\star - G_{nn}|c_n|^2 \mright\} \mright \} \label{eq:f_fn} \\
    I_{nm}(c_n,c_m) &:= \exp \mleft \{ -\frac{2}{\sigma^2} {\text{Re}} \mleft\{  G_{nm} c_m c_n^\star  \mright\} \mright \}. \label{eq:I_fn}
\end{align}
\begin{figure}[tb] %
\centering
\tikzstyle{fn} = [draw, very thick, regular polygon, regular polygon sides=4, minimum width = 2.5em, inner sep=0pt, rounded corners]
\tikzstyle{vn} = [draw, very thick, circle, inner sep=0pt, minimum size = 2em]
\begin{tikzpicture}[auto, node distance=3em and 3.5em, thick]
\clip (1.7, -2.1) rectangle (9.5, 0.37);
    \node [vn, draw=none] (c0){};
    \node [vn, draw=none, right= of c0] (c1){};
    \node [vn, label=center:$c_{3}$, right= of c1] (c2){};
    \node [vn, label=center:$c_{4}$, right= of c2] (c3){};
    \node [vn, label=center:$c_{5}$, right= of c3] (c4){};
    \node [vn, draw=none, right= of c4] (c5){};
    \node [fn, draw=none, right=0.5em of c4] (rdots) {$\cdots$};
    \node [vn, draw=none, right= of c5] (c6){};
    \node [fill=KITcyan!35!white, fn, label=center: $F_{3}$, left=0.5em of c2] (p2){};
    \node [fn, draw=none, left=0.5em of p2] (rdots) {$\cdots$};
    \node [fill=KITcyan!35!white, fn, label=center: $F_{4}$, left=0.5em of c3] (p3){};
    \node [fill=KITcyan!35!white, fn, label=center: $F_{5}$, left=0.5em of c4] (p4){};
    \node [fill=KITorange!50!white, fn, label=center: $I_{31}$, below= of c2.260, anchor = north east] (I20){};
    \node [fill=KITorange!50!white, fn, label=center: $I_{32}$, below= of c2.280, anchor = north west] (I21){};
    \node [fill=KITorange!50!white, fn, label=center: $I_{42}$, below= of c3.260, anchor = north east] (I31){};
    \node [fill=KITorange!50!white, fn, label=center: $I_{43}$, below= of c3.280, anchor = north west] (I32){};
    \node [fill=KITorange!50!white, fn, label=center: $I_{53}$, below= of c4.260, anchor = north east] (I42){};
    \node [fill=KITorange!50!white, fn, label=center: $I_{54}$, below= of c4.280, anchor = north west] (I43){};
    \node [fn, draw=none, below= of c5.260, anchor = north east] (I53){};
    \node [fn, draw=none, below= of c5.280, anchor = north west] (I54){};
    \node [fn, draw=none, below= of c6.260, anchor = north east] (I64){};
    \node [fn, draw=none, below= of c6.280, anchor = north west] (I65){};
    \draw[-] (c2.west) -- (p2.east);
    \draw[-] (c3.west) -- (p3.east);
    \draw[-] (c4.west) -- (p4.east);
    \draw[-] (I20.north) -- (c2.south);
    \draw[dotted] (I20.north) -- ($(I20.north)!0.4!(c0.south)$);
    \draw[-] (I20.north) -- ($(I20.north)!0.2!(c0.south)$);
    \draw[-] (I21.north) -- (c2.south);
    \draw[-] (I21.north) -- ($(I21.north)!0.58!(c1.south)$);
    \draw[dotted] (I21.north) -- ($(I21.north)!1.0!(c1.south)$);
   \draw[-] (c2.south) -- (I32.north);
    \draw[-] (c2.south) -- (I42.north);
    \draw[-] (I31.north) -- (c3.south);
    \draw[-] (I31.north) -- ($(I31.north)!0.7!(c1.south)$);
    \draw[dotted] (I31.north) -- ($(I31.north)!1.0!(c1.south)$);
    \draw[-] (I32.north) -- (c3.south);
    \draw[-] (c3.south) -- (I43.north);
    \draw[-] (c3.south) -- ($(c3.south)!0.8!(I53.north)$);
    \draw[dotted] (c3.south) -- ($(c3.south)!1.0!(I53.north)$);
    \draw[-] (I42.north) -- (c4.south);
    \draw[-] (I43.north) -- (c4.south);
    \draw[-] (c4.south) -- ($(c4.south)!0.4!(I54.north)$);
    \draw[dotted] (c4.south) -- ($(c4.south)!0.8!(I54.north)$);
    \draw[-] (c4.south) -- ($(c4.south)!0.28!(I64.north)$);
    \draw[dotted] (c4.south) -- ($(c4.south)!0.4!(I64.north)$);
\end{tikzpicture}
\caption{Factor graph representation of \eqref{eq:Ungerboeck_factorization} for a channel with memory $L=2$.}
    \label{fig:factor_graph_ungerboeck}
\end{figure}
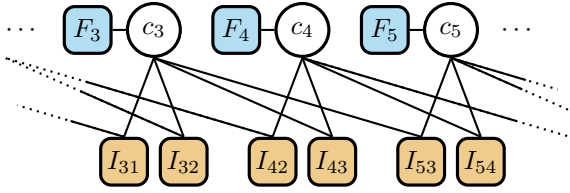

The factor graph representation of \eqref{eq:Ungerboeck_factorization} is shown in Fig.~\ref{fig:factor_graph_ungerboeck}.
Applying the \ac{BP} algorithm with a parallel schedule on the factor graph in Fig.~\ref{fig:factor_graph_ungerboeck} yields an iterative algorithm for symbol detection. We denote by ${\mu^{(t)}_{nm}(c_m)}$ a \ac{BP} message from variable $c_n$ to $c_m$ in iteration~$t$.
After a uniform initialization of all messages~${\mu^{(0)}_{nm}(c_m)=1}$, the iterative \ac{BP} updates read for ${t=1,\ldots,T}$ as
\begin{align}
    b_n^{(t)}(c_n) &= \gamma_n^{(t)} \prod\limits_{k \neq n} \mu_{kn}^{(t)}(c_n), \label{eq:belief_update} \\
    \mu_{nm}^{(t+1)}(c_m) &= \sum\limits_{c_n} I_{nm}(c_n,c_m) F_n(c_n) \frac{b_n^{(t)}(c_n)}{\mu_{mn}^{(t)}(c_n)}, \label{eq:message_update}
\end{align}
where ${\gamma_{n}^{(t)} \in \mathbb{R}}$ is a normalization constant such that ${b^{(t)}_n(c_n)}$ is a normalized probability distribution.\footnote{Note that in practical implementations, the \ac{BP} algorithm is typically carried out in the logarithmic domain for improved numerical stability and reduced computational complexity.}
Because the factor graph in Fig.~\ref{fig:factor_graph_ungerboeck} contains cycles, the beliefs~${b_n(c_n)}$ are only an approximation of the \ac{APP} distributions~${P(c_n|\bm{y})}$ and the quality of approximation strongly depends on the underlying channel characteristics~\cite{schmid_low-complexity_2022}.

\section{Blind Estimation and Detection via EM and BP} \label{sec:EMBP}
The detection algorithm in Sec.~\ref{sec:ufg} relies on the availability of \ac{CSI} in the definition of the factors~\eqref{eq:f_fn} and~\eqref{eq:I_fn}. To enable blind detection, we propose an iterative joint estimation and detection scheme by integrating the \ac{EM} algorithm into the iterative receiver.
 \ac{EM}  is an established method for finding \ac{ML} parameter estimates in probabilistic models that contain latent variables~\cite{dempster_maximum_1977}.
The log-likelihood function can written as
\begin{align*}
    &\log p(\bm{y}|\bm{\theta}) = \log p(\bm{c},\bm{y}|\bm{\theta}) - \log P(\bm{c}|\bm{y},\bm{\theta}) \nonumber \\
    &= \underbrace{\sum\limits_{\bm{c}} Q(\bm{c}) \log \mleft( \frac{p(\bm{c},\bm{y}|\bm{\theta})}{Q(\bm{c})} \mright)}_{=: \mathcal{L}(Q,\bm{\theta})} \underbrace{- \sum\limits_{\bm{c}} Q(\bm{c}) \log \mleft( \frac{P(\bm{c}|\bm{y},\bm{\theta})}{Q(\bm{c})} \mright)}_{=: D_\text{KL}(Q \Vert P) \geq 0}, \label{eq:likelihood_decomposition}
\end{align*}
where $Q(\bm{c})$ is a normalized trial distribution of $\bm{c}$.
The \ac{EM} algorithm maximizes the \ac{ELBO} ${\mathcal{L}(Q,\bm{\theta})} \leq { \log p(\bm{y}|\bm{\theta})}$ in a 2-stage iterative algorithm.
Starting with an initial guess for the parameters~$\hat{\bm{\theta}}^{(0)}$, the \ac{EM} algorithm iteratively performs for ${t=1,\ldots,T}$
\begin{enumerate}
    \item \textbf{E-step}: Evaluate 
    \begin{equation}
        {Q^{(t)}(\bm{c})=P(\bm{c}|\bm{y},\hat{\bm{\theta}}^{(t-1)})}, \label{eq:EM_Mstep}
    \end{equation}
    \item \textbf{M-step}: Compute 
    \begin{equation}
        {\hat{\bm{\theta}}^{(t)}} = \arg\max\limits_{\bm{\theta}} \mathcal{L}\mleft( Q^{(t)},\bm{\theta} \mright). \label{eq:EM_Estep}
    \end{equation}
\end{enumerate}
The \ac{EM} algorithm monotonically increases the log-likelihood~${\log p(\bm{y}|\hat{\bm{\theta}}^{(t)})}$ in every iteration~$t$, however, it is not guaranteed to converge to the global maximum~\cite{eckford_channel_2004}.

\begin{figure*}
\begin{align}
&\mspace{0mu}\hat{\sigma}^{2(t)} = \frac{1}{N} \Bigg[\sum\limits_{n=1}^{N+L} |y_n|^2 - \sum\limits_{n=1}^N \sum\limits_{c_n} b^{(t)}_n(c_n) \bigg( 2 \text{Re} \left\{ x_n c_n^\star \right\} - G_{nn} |c_n|^2 - \sum\limits_{m<n} \sum\limits_{c_m} b^{(t)}_m(c_m) 2 \text{Re} \left\{ G_{mn} c_m c_n^\star \right\} \bigg) \Bigg], \label{app:eq:sigma2} \\
&\begin{aligned}
     \hat{h}_{\ell}^{(t)} 
     =  \Bigg( &\sum\limits_{n=1}^N \sum\limits_{c_n} b^{(t)}_n(c_n)  (y_{n+\ell} c_n^\star) - \sum\limits_{\substack{k=0 \\ k \neq \ell}}^L \sum\limits_{c_{n-|\ell-k|}} b^{(t)}_{n-|\ell-k|}(c_{n-|\ell-k|}) h^{(t-1)}_{k} \cdot  \\
     &\bigg( \text{Re}\{ c_{n-|\ell-k|} c_n^\star \} 
     -\text{j} \cdot \text{Im}\{ c_{n-|\ell-k|} c_n^\star \} \text{sign} \mleft\{ \ell-k \mright\} \bigg) \Bigg) 
     \cdot \left( \sum\limits_{n=1}^N \sum\limits_{c_n} b^{(t)}_n(c_n) |c_n|^2 \right)^{-1}
\end{aligned} \label{app:eq:update_h_complex}
\end{align}
\hrule
\end{figure*}
We apply the \ac{EM} algorithm to the considered problem in Sec.~\ref{subsec:system_model}. 
The \Estep\ in~\eqref{eq:EM_Estep} requires the computation of the \ac{APP} distribution ${Q^{(t)}(\bm{c})=P(\bm{c}|\bm{y},\hat{\bm{\theta}}^{(t-1)})}$, which relates to the \ac{MAP} symbol detection task discussed in Sec.~\ref{sec:ufg}. To reduce the computational complexity, we consequently use the \ac{BP}-based symbol detector to efficiently perform the \Estep\ by approximating the symbol-wise posteriors $P(c_n|\bm{y},\hat{\bm{\theta}}^{(t-1)})$ with the beliefs $b_n(c_n)$. Thereby, the complexity only scales linearly with $L$, instead of exponentially.
For the optimization~\eqref{eq:EM_Estep} in the \Mstep, we follow a coordinate descent maximization. By solving ${(\partial/ {\partial \theta_i}) \mathcal{L}\mleft(b_n(c_n)^{(t)},\bm{\theta} \mright)  = 0}$ for every~$\theta_i$, we can find the optimal update rules for each dimension $\theta_i$ in closed form, respectively. The resulting \ac{EM} parameter updates are given in ~\eqref{app:eq:sigma2}-\eqref{app:eq:update_h_complex}, and scale with ${\mathcal{O}\big( NLM^2 \big)}$ and ${\mathcal{O}\big( NL^2M \big)}$, respectively. A detailed derivation is provided in the full version of this paper~\cite{schmid2024blind}.

\subsection{Optimizing the \ac{EM} Parameter Update Schedule} \label{sec:schedule}
The \ac{EM} parameter updates directly depend on each other, e.g., the update of~$\hat{h}_\ell^{(t)}$ in \eqref{app:eq:update_h_complex} is a function of $\hat{h}_k^{(t-1)}$ with ${k\neq \ell}$.
This raises the question of a suitable schedule, i.e., which parameters of~$\bm{\theta}$ to update simultaneously in which iteration~$t$ of the \ac{EM} algorithm.
Learning a schedule is a discrete and combinatorial problem that is in general non-trivial. We simplify the problem by first converting the schedule learning task into a continuous optimization problem which is then combined with a simple pruning strategy.
Therefore, we replace the original \ac{EM} parameter estimates $\hat{\theta}^{(t)}_k$ by a convex combination of old and new estimates ${\beta_{\text{EM},k}^{(t)} \hat{{\theta}}^{(t)}_k} + {({1}-{\beta}_{\text{EM},k}^{(t)}) \hat{\theta}_k^{(t-1)}}$ where we have introduced the continuous weights ${(\beta_{\text{EM},1}^{(t)},\ldots,\beta_{\text{EM},L+2}^{(t)})^{\rm T} =: \bm{\beta}_\text{EM}^{(t)} \in \mathbb{R}^{L+2}},\, {t=1,\ldots,T}$. Note that ${\beta_{\text{EM},k}^{(t)}=1}$ corresponds to a full update of the parameter estimate~$\hat{\theta}_k$, and ${\beta_{\text{EM},k}^{(t)}=0}$ means that this parameter is not updated at all in the \ac{EM} step~$t$.

We optimize the weights ${\bm{\beta}_\text{EM}^{(t)}}$ based on a training data set
\begin{equation*}
    \label{eqn:DataSet}
    {\mathcal{D}=\{ (\bm{c}^{(d)},\bm{\theta}^{(d)},\bm{y}^{(d)}): \, d=1,\ldots,D \}},
\end{equation*}
which consists of randomly and independently sampled information sequences $\bm{c}^{(d)}$, channel realizations $\bm{\theta}^{(d)}$, 
and the corresponding noisy channel observations $\bm{y}^{(d)}$.
By unrolling the \ac{EM} iterations~$t$ into a trainable discriminative machine learning architecture~\cite{shlezinger2022discriminative}, we tune the weights $\bm{\beta}_\text{EM}^{(t)}$ based on~$\mathcal{D}$ using backpropagation and the   Adam algorithm~\cite{kingma_adam_2015}.

Objective functions for end-to-end optimization can either focus on the estimation or the detection performance.
For the latter, we use the \ac{BMI} as a maximum achievable information rate between the channel input and the output of the symbol detector~\cite{alvarado_achievable_2018}.
As a loss function for the channel estimation task, we consider the squared estimation error
\begin{equation*}
    \lVert \hat{\bm{h}} - \bm{h} \rVert^2 = \sum_{\ell=0}^L \big| \hat{h}_\ell-h_\ell \big|^2
\end{equation*}
of the channel impulse response~$\bm{h}$.

To limit the overall number of parameter updates $K_\text{EM} < T \cdot \left( L+2 \right)$, we add the $L_1$~loss 
\begin{equation*}
    \mathcal{L}_\text{EM}(K') := \sum\limits_{\beta_{k} \in \mathcal{\beta}_{\text{min}}(K')} \lvert \beta_{k} \rvert
\end{equation*}
that is known to encourage the sparsity of the coefficients to the original loss function. We denote by~$\mathcal{\beta}_{\text{min}}(K')$ the set of $K'$ scalar elements in the vectors $\bm{\beta}_\text{EM}^{(t)},\, t=1,\ldots,T$ with the smallest absolute value. Thereby, the regularization penalizes more than ${T\cdot\left( L+2 \right) - K'}$ parameter updates by enforcing $K'$ scalar elements of the vectors $\bm{\beta}_\text{EM}^{(t)},\, \forall t$ to be close to zero, i.e., they are effectively not  updated.
During training, we initialize ${K'=0}$ and gradually increase $K'$ until the target number of parameter updates ${K_\text{EM} = T\cdot\left( L+2 \right) - K'}$ is reached. After the training procedure, we implement the learned schedule by setting the $K'$ elements in ${\mathcal{\beta}_{\text{min}}(K')}$ to zero. This effectively reduces the  complexity since $K'$ less parameter update computations via~\eqref{app:eq:sigma2} or \eqref{app:eq:update_h_complex} need to be carried out.
For the remaining parameter updates, we apply the learned momentum weights $\bm{\beta}_\text{EM}^{(t)}$ as a simple and low-complexity generalization of the full \ac{EM} updates~(${\bm{\beta}_\text{EM}^{(t)}=\bm{1}}$). 

\subsection{Enhancing \ac{BP} Accuracy} \label{subsubsec:momentumBP}
The concept of momentum can also be applied to the \ac{BP} algorithm. By only partially updating the \ac{BP} messages in each iteration, the convergence behavior can be significantly improved, while retaining the same \ac{BP} fixed points~\cite{murphy_loopy_1999}.
As opposed to Sec.~\ref{sec:schedule}, we apply only one momentum weight ${\beta_\text{BP}^{(t)} \in \mathbb{R}}$ per \ac{BP} message passing iteration~${t=1,\ldots,T}$.
Compared to other neural enhancement techniques like \emph{neural \ac{BP}} that assign an individual weight to every edge in the unrolled factor graph~\cite{nachmani_deep_2018, schmid_low-complexity_2022}, we expect our model to better generalize to a broad range of channel characteristics instead of specializing in one specific channel.

\begin{algorithm}[t] %
    \DontPrintSemicolon
    \KwData{$\bm{y}$, $\hat{\bm{\theta}}^{(0)}$} %
    ${\mu^{(0)}_{nm}(x_m)=1, b_n^{(0)}(x_n)}, \quad {n,m=1,\ldots N}, \, n \neq m$ \\
    \For{$t = 1,\ldots,T$}
    {
        \textbf{E-step}: \ac{BP} message passing \\
        Compute messages $\mu_{nm}^{(t)}(c_m)$ using~\eqref{eq:message_update} \\
        $\mu_{nm}^{(t)}(c_m) \gets \beta_\text{BP}^{(t)} \mu_{nm}^{(t)}(c_m) + (1-\beta_\text{BP}^{(t)}) \mu_{nm}^{(t-1)}(c_m)$ \\
        Compute beliefs~$b^{(t)}_n(c_n)$ based on~\eqref{eq:belief_update}\\
        \textbf{M-step}: Update parameter estimates \\
        Compute $\hat{\bm{\theta}}^{(t)}$ using~\eqref{app:eq:sigma2} and \eqref{app:eq:update_h_complex}  \\
        $\hat{\bm{\theta}}^{(t)} \gets \bm{\beta}_\text{EM}^{(t)} \odot \hat{\bm{\theta}}^{(t)} + (\bm{1}-\bm{\beta}_\text{EM}^{(t)}) \odot \hat{\bm{\theta}}^{(t-1)}$
    }
    \KwResult{${\hat{\bm{\theta}}=\hat{\bm{\theta}}^{(T)}}, \; {b_n(c_n) = b_n^{(T)}(c_n)}$}
    \caption{EMBP} %
    \label{alg:EMBP}
\end{algorithm}
\subsection{The EMBP Algorithm}
We summarize the proposed blind estimation and joint detection scheme in Alg.~\ref{alg:EMBP}. Since it combines message passing iterations from the \ac{BP} algorithm with \ac{EM} parameter update steps, we denote it the \emph{EMBP algorithm}.
For the required initialization of the parameters~$\hat{\bm{\theta}}^{(0)}$, we found that the \ac{VAELE}, a lightweight blind channel equalizer~\cite{lauinger_blind_2022}, can provide a robust and convenient initialization.

\section{Simulation Study} \label{sec:experiments}
The numerical simulations in this section consider the transmission blocks to contain~${N=100}$ symbols from a \ac{BPSK} constellation. 
For initialization of the EMBP algorithm, we apply $10$ iterations of the \ac{VAELE} algorithm~\cite{lauinger_blind_2022} with a linear equalization filter of length ${2L+1}$. 

\begin{figure}
    \centering
    \begin{tikzpicture}
    \pgfplotsset{
        legend image dotted/.style={
            legend image code/.code={%
            \draw[dashdotted, color=black!100!white] (0cm,0.05cm) -- (0.3cm,0.05cm);
            \draw[dashdotted, color=black!60!white] (0cm, -0.05cm) -- (0.3cm, -0.05cm);
            }
        },
    }
    \pgfplotsset{
legend image code/.code={
\draw[mark repeat=2,mark phase=2]
plot coordinates {
(0cm,0cm)
(0.15cm,0cm)        %
(0.3cm,0cm)         %
};%
}
}
    \begin{axis}[
    width=\linewidth, %
    height=0.75\linewidth,
    align = left,
    grid=major, %
    grid style={gray!30}, %
    xlabel= $\mathsf{snr}$ (dB),
    ylabel= BER,
	  scaled y ticks=false,
    ymode = log,
    ymin = 0.0003,
    ymax = 0.5,
    xmin = -4,
    xmax = 12,
    enlarge x limits=false,
    enlarge y limits=false,
    line width=1pt,
	  legend style={font=\footnotesize, cells={align=left}, anchor=south west, at={(0.02,0.02)}},
    legend cell align={left},
	  smooth,
    ]
    \addplot[color=VAEcolor, line width=1pt] table[x={EsN0 dB}, y={VAE BER}, col sep=comma] {numerical_results/BER+MSE_over_EsN0.csv};
    \addlegendentry{VAE-LE}
    \addplot[color=EMBPcolor, line width=1pt] table[x={EsN0 dB}, y={EMBP(VAE) BER}, col sep=comma] {numerical_results/BER+MSE_over_EsN0.csv};
    \addlegendentry{EMBP, ${\beta_\text{BP}=1}$}
    \addplot[color=EMNBPcolor, line width=1pt] table[x={EsN0 dB}, y={EMNBP(VAE) BER}, col sep=comma] {numerical_results/BER+MSE_over_EsN0.csv};
    \addlegendentry{EMBP, $\beta_\text{BP}^\star$}
    \addplot[color=BPcolor, line width=1pt, dashed] table[x={EsN0 dB}, y={BP(hEMBP) BER}, col sep=comma] {numerical_results/BER+MSE_over_EsN0.csv};
    \addlegendentry{BP, ${\hat{\bm{h}}_\text{EMBP}}$}
    \addplot[color=BPcolor, line width=1pt, densely dotted] table[x={EsN0 dB}, y={BP(genie) BER}, col sep=comma] {numerical_results/BER+MSE_over_EsN0.csv};
    \addlegendentry{BP, $\bm{h}$ (coherent)}
    \addplot[color=black!30!white, line width=1pt, densely dotted] table[x={Es/N0 (dB)}, y={genie MAP BER}, col sep=comma] {numerical_results/results_genie_MAP.csv};
    \addlegendentry{MAP, $\bm{h}$ (coherent)}
    \addlegendimage{legend image dotted}
    \addplot[color=black!100!white, line width=1pt, dashdotted] table[x={Es/N0 (dB)}, y={BER MAP pilots}, col sep=comma] {numerical_results/results_5pc_pilots.csv};
    \addlegendentry{MAP, $\hat{\bm{h}}_\text{ML}$, \\ $5/10\%$ pilots}
    \addplot[color=black!60!white, line width=1pt, dashdotted] table[x={Es/N0 (dB)}, y={BER MAP pilots}, col sep=comma] {numerical_results/results_10pc_pilots.csv};
    \node at (axis cs:10,0.02) {\footnotesize \color{black!100!white} ${5\%}$};
    \node at (axis cs:10,0.0012) {\footnotesize \color{black!60!white} ${10\%}$};
  \end{axis}
\end{tikzpicture}
    \caption{BER over $\mathsf{snr}$ for various detection schemes, averaged over $10^7$ random channels with ${L=2}$. For the EMBP algorithm, a fixed serial \ac{EM} update schedule is applied.}
    \label{fig:BER_vs_SNR}
\end{figure}
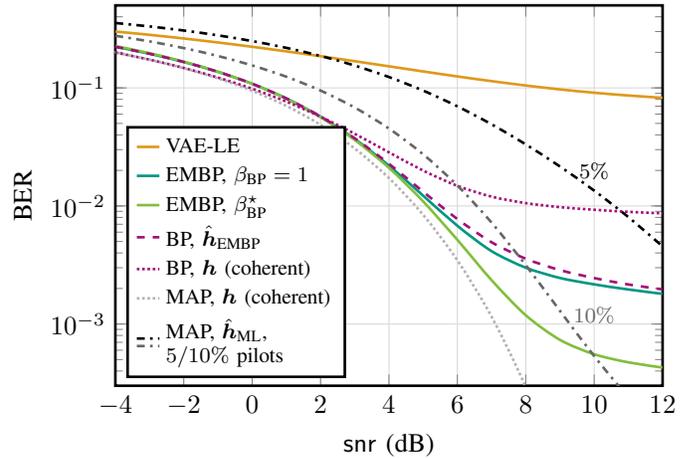

We evaluate the detection performance of the EMBP algorithm in terms of the \ac{BER} over the \ac{SNR} in Fig.~\ref{fig:BER_vs_SNR} for $10^7$ random channels with memory ${L=2}$. Note that despite the relatively short channel impulse response length~${L+1=3}$, the sampled channels can introduce strong \ac{ISI} where linear detectors like the \ac{FIR} filter of the \ac{VAELE} have limited performance~\cite{proakis_digital_2007}.
This can also be observed in Fig.~\ref{fig:BER_vs_SNR}, where the \ac{VAELE} is significantly outperformed by the EMBP detector. For the latter, we do not yet apply the momentum-based optimization techniques introduced in Sec.~\ref{sec:schedule} and~\ref{subsubsec:momentumBP}. Instead, we use a serial parameter update schedule, i.e., every \ac{EM} step only updates one parameter. We use ${T=3\cdot \left( L+2 \right)}$ iteration which means that every parameter $\theta_k$ is updated $3$~times. We also disable the momentum in the \ac{BP} message passing by fixing ${\beta^{(t)}_\text{BP}=1}$ for all $t$.
A surprising observation in Fig.~\ref{fig:BER_vs_SNR} is that the blind EMBP algorithm clearly outperforms the coherent \ac{BP} detector for ${\mathsf{snr} > 3}$~dB. 
At first glance, it might seem counterintuitive that a blind detector can outperform its coherent counterpart which has full access to the \ac{CSI}. However, the suboptimality of the \ac{BP} algorithm on cyclic factor graphs directly and heavily depends on the characteristics of the underlying channel assumption.
In the full version of this paper~\cite{schmid2024blind}, we could demonstrate that it is an inherent property of the \ac{EM} algorithm to converge to a ``surrogate'' channel representation that deviates from the true channel such that it is better suited for \ac{BP}-based detection~\cite{schmid2024blind}.
This is also confirmed in Fig.~\ref{fig:BER_vs_SNR}, where the original \ac{BP} detector based on the channel estimate~$\hat{\bm{h}}_\text{EMBP}$ of the EMBP algorithm performs similarly to the EMBP algorithm itself.

We compare the results of the blind EMBP detector to \ac{MAP} detection based on a pilot-aided channel estimation. To this end, we fix a portion of the $N$ information symbols in the transmission block to pseudo-random pilot symbols that are known at the receiver. Based on these pilot symbols, the receiver performs \ac{ML} estimation of~$\bm{h}$.
Figure~\ref{fig:BER_vs_SNR} shows that the EMBP algorithm clearly outperforms the \ac{MAP} detector where $5\%$ of the transmission symbols are pilot symbols. For a target \ac{BER} of $10^{-2}$, the EMBP detector also outperforms the baseline with $10\%$ pilots with a $1$~dB gain. In the high \ac{SNR} regime, we observe that the dominant source of error for the EMBP algorithm is the suboptimality of the \ac{BP} detector. Many short cycles in the underlying factor graph can cause a non-convergent behavior of the \ac{BP} algorithm, especially for high \ac{SNR}~\cite{schmid_low-complexity_2022}.
To reduce this error floor, we can leverage the momentum-based \ac{BP} message updates as introduced in Sec.~\ref{subsubsec:momentumBP}. We optimize the weights ${\beta_\text{BP}^{(t)}, t=1,\ldots,L+2=12}$ for generic channels, i.e., each sample in the training data set $\mathcal{D}$ consists of an independent channel realization and the ${\mathsf{snr}}$ is uniformly sampled in $[0\,\text{dB},12\,\text{dB}]$. This generic training can be performed offline and prior to evaluation.
We apply batch gradient descent optimization with $200$ batches each containing $1000$ transmission blocks. 
As shown in Fig.~\ref{fig:BER_vs_SNR}, the EMBP algorithm with optimized momentum weights~$\beta_\text{BP}^\star$ can effectively reduce the error floor by almost one order of magnitude.

We further study the capability of the proposed method to learn an effective \ac{EM} update schedule that reduces the computational complexity of the EMBP algorithm. We consider a transmission over randomly sampled channels with memory ${L=5}$ and optimize the EMBP algorithm towards minimum \ac{MSE} of the EMBP channel estimates~$\hat{\bm{h}}$ while restraining the algorithm's complexity. More specifically, we limit the \ac{EM} iterations to ${T=6}$ or ${T=3}$, respectively. Based on this restriction ${T < L+2=7}$, the EMBP algorithm would not be able to update each parameter (including $\sigma^2$) at least once if it followed a serial update schedule where only one parameter is updated per \ac{EM} step. We apply the gradient-based end-to-end training method detailed in Sec.~\ref{sec:schedule} to find a more effective \ac{EM} update schedule for this setting.
We use $2500$ batches for offline training and each batch contains $1000$ random channel realizations. For the schedule learning, we fix the maximum number of parameter updates to ${K_\text{EM}=24}$. 

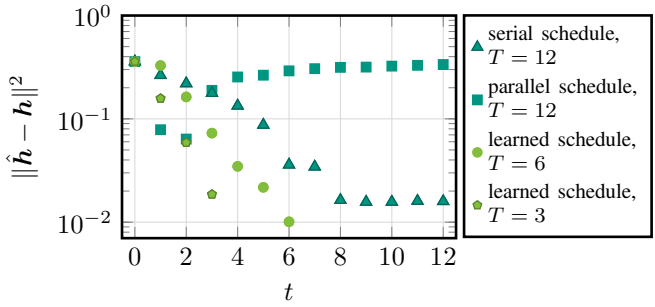
\begin{figure}
    \centering
    \begin{tikzpicture}
    \begin{axis}[
    ymode=log,
    width=0.68\linewidth, %
    height= 0.513\linewidth,
    align = left,
    grid=major, %
    grid style={gray!30}, %
    xlabel= $t$,
    ylabel= ${\lVert \hat{\boldsymbol{h}} - \boldsymbol{h} \rVert^2}$,
    ymin = 0.007,
    ymax = 1.0,
    xmin = -0.5,
    xmax = 12.5,
    xtick = {0,2,4,6,8,10,12},
    enlarge x limits=false,
    enlarge y limits=false,
    line width=1pt,
    legend style={font=\footnotesize, cells={align=left}, anchor=south west, at={(1.02, 0.0)}},
    legend cell align={left},
    ]
    \addplot[draw=none, color=EMBPcolor, draw=EMBPcolor!70!black, mark=triangle*, line width=0.5pt, only marks, mark options={scale=1.2}] table[x={iters}, y={EMBPserial SE h}, col sep=comma] {numerical_results/eval_over_iters.csv};
    \addlegendentry{serial schedule,\\ ${T=12}$}
    \addplot[color=EMBPcolor, mark=square*, line width=0.5pt, only marks, mark options={scale=0.9}] table[x={iters}, y={EMBPparallel SE h}, col sep=comma] {numerical_results/eval_over_iters.csv};
    \addlegendentry{parallel schedule,\\ ${T=12}$}
    \addplot[draw=none, color=EMNBPcolor, mark=*, only marks, mark options={scale=0.75}] table[x={iters}, y={EMBPopt6 SE h}, col sep=comma] {numerical_results/eval_over_iters.csv};
    \addlegendentry{learned schedule,\\${T=6}$}
    \addplot[draw=none, color=EMNBPcolor, draw=EMNBPcolor!70!black, mark=pentagon*, line width=0.5pt, only marks, mark options={scale=0.85}] table[x={iters}, y={EMBPopt3 SE h}, col sep=comma] {numerical_results/eval_over_iters.csv};
    \addlegendentry{learned schedule,\\${T=3}$}
    \addplot[color=EMBPcolor, draw=EMBPcolor!70!black, mark=triangle*, line width=0.5pt, only marks, mark options={scale=1.2}] table[x={iters}, y={EMBPserial SE h}, col sep=comma] {numerical_results/eval_over_iters.csv};
    \addplot[color=EMNBPcolor, mark=*, only marks, mark options={scale=0.75}] table[x={iters}, y={EMBPopt6 SE h}, col sep=comma] {numerical_results/eval_over_iters.csv};
    \addplot[color=EMNBPcolor, draw=EMNBPcolor!70!black, mark=pentagon*, line width=0.5pt, only marks, mark options={scale=0.85}] table[x={iters}, y={EMBPopt3 SE h}, col sep=comma] {numerical_results/eval_over_iters.csv};
    \end{axis}
    \end{tikzpicture}
    \caption{Mean squared estimation error ${\lVert \hat{\boldsymbol{h}} - \boldsymbol{h} \rVert^2}$ after each iteration of the EMBP algorithm with different \ac{EM} parameter update schedules. The randomly sampled channels have ${\mathsf{snr}=10}$~dB and memory ${L=5}$, i.e., $7$~parameters to estimate (including $\sigma^2$).}
    \label{fig:SE_over_iters}
\end{figure}

Figure~\ref{fig:SE_over_iters} evaluates the squared channel estimation error over the EMBP iterations~$t$ for different \ac{EM} update schedules. The results are averaged over $10^5$ random channels with ${\mathsf{snr}=10}$~dB.
For the serial schedule, the \ac{MSE} monotonically decreases with the EMBP iterations~$t$.
The stagnation at ${t=7}$ is due to the parameter estimate $\hat{\sigma}^2$ not being \emph{directly} included in the evaluation of the squared error ${\lVert \hat{\boldsymbol{h}} - \boldsymbol{h} \rVert^2}$.
The second baseline in Fig.~\ref{fig:SE_over_iters} uses a parallel schedule for   EMBP, i.e., all parameters are simultaneously updated in every \ac{EM} step~$t$. As a result, the \ac{MSE} decreases significantly after the first \ac{EM} iteration but worsens after ${t=2}$ due to an unstable behavior of the fully parallel updates.
The EMBP algorithm with optimized update schedules shows a fast yet stable convergence and the estimation results are comparable to those of the serial \ac{EM} schedule but with less iterations, i.e., with a reduced computational complexity.
The analysis of the learned schedules revealed that only a few parameters are partially updated with small weights~$\beta_\text{EM}$ in the early \ac{EM} iterations while the last iterations contain full and parallel updates.

\section{Conclusion}
We proposed the EMBP algorithm for blind channel estimation and joint symbol detection. Combining two iterative inference schemes, \ac{EM} for estimation and \ac{BP} for detection, yields a joint estimation and detection scheme with superior performance and low complexity. It further poses the challenge of optimizing the interaction between the iterations of both algorithms. For the EMBP algorithm, we found the concept of momentum as a particularly simple and effective method to improve an iterative model while retaining an algorithm that generalizes well.

\bibliographystyle{IEEEtran}
\bibliography{IEEEabrv,bibliography}

\begin{thebibliography}{10}
\providecommand{\url}[1]{#1}
\csname url@samestyle\endcsname
\providecommand{\newblock}{\relax}
\providecommand{\bibinfo}[2]{#2}
\providecommand{\BIBentrySTDinterwordspacing}{\spaceskip=0pt\relax}
\providecommand{\BIBentryALTinterwordstretchfactor}{4}
\providecommand{\BIBentryALTinterwordspacing}{\spaceskip=\fontdimen2\font plus
\BIBentryALTinterwordstretchfactor\fontdimen3\font minus
  \fontdimen4\font\relax}
\providecommand{\BIBforeignlanguage}[2]{{%
\expandafter\ifx\csname l@#1\endcsname\relax
\typeout{** WARNING: IEEEtran.bst: No hyphenation pattern has been}%
\typeout{** loaded for the language `#1'. Using the pattern for}%
\typeout{** the default language instead.}%
\else
\language=\csname l@#1\endcsname
\fi
#2}}
\providecommand{\BIBdecl}{\relax}
\BIBdecl

\bibitem{ghosh_maximum-likelihood_1992}
M.~Ghosh and C.~L. Weber, ``Maximum-likelihood blind equalization,''
  \emph{Optical Engineering}, vol.~31, no.~6, pp. 1224--1228, 1992.

\bibitem{godard_self-recovering_1980}
D.~Godard, ``\BIBforeignlanguage{en}{Self-{recovering} {equalization} and
  {carrier} {tracking} in {two}-{dimensional} {data} {communication}
  {systems}},'' \emph{\BIBforeignlanguage{en}{{IEEE} Trans. Commun.}}, vol.~28,
  no.~11, pp. 1867--1875, 1980.

\bibitem{lauinger_blind_2022}
V.~Lauinger, F.~Buchali, and L.~Schmalen, ``Blind {equalization} and {channel}
  {estimation} in {coherent} {optical} {communications} {using} {variational}
  {autoencoders},'' \emph{IEEE J. Sel. Areas Commun.}, vol.~40, no.~9, pp.
  2529--2539, Sep. 2022.

\bibitem{worthen_unified_2001}
A.~Worthen and W.~Stark, ``Unified design of iterative receivers using factor
  graphs,'' \emph{IEEE Transactions on Information Theory}, vol.~47, no.~2, pp.
  843--849, Feb. 2001.

\bibitem{loeliger_remarks_2003}
H.-A. Loeliger, ``Some remarks on factor graphs,'' in \emph{Proc. {Int}.
  {Symposium} {Turbo} {Codes} and {Related} {Topics} ({ISTC})}, Brest, France,
  Sep. 2003, pp. 111--115.

\bibitem{liu_joint_2009}
Y.~Liu, L.~Brunel, and J.~J. Boutros, ``\BIBforeignlanguage{en}{Joint channel
  estimation and decoding using {Gaussian} approximation in a factor graph over
  multipath channel},'' in \emph{\BIBforeignlanguage{en}{Proc. IEEE Int. Symp.
  Pers. Indoor Mobile Radio Commun. (PIMRC)}}, Tokyo, Japan, Sep. 2009, pp.
  3164--3168.

\bibitem{eckford_channel_2004}
A.~W. Eckford, ``Channel {Estimation} in {Block} {Fading} {Channels} {Using}
  the {Factor} {Graph} {EM} {Algorithm},'' in \emph{Proc. Biennial Symposium on
  Communications}, Kingston, Ontario, Canada, May 2004.

\bibitem{dauwels_expectation_2005}
J.~Dauwels, S.~Korl, and H.-A. Loeliger, ``Expectation maximization as message
  passing,'' in \emph{Proc. IEEE Int. Symp. Inf. Theory (ISIT)}, Adelaide,
  Australia, Sep. 2005, pp. 583--586.

\bibitem{shlezinger_model-based_2023}
N.~Shlezinger, J.~Whang, Y.~C. Eldar, and A.~G. Dimakis, ``Model-{based} {deep}
  {learning},'' \emph{Proc. {IEEE}}, vol. 111, no.~5, pp. 465--499, May 2023.

\bibitem{shlezinger2022model}
N.~Shlezinger, Y.~C. Eldar, and S.~P. Boyd, ``Model-based deep learning: On the
  intersection of deep learning and optimization,'' \emph{{IEEE} Access},
  vol.~10, pp. 115\,384--115\,398, 2022.

\bibitem{proakis_digital_2007}
J.~Proakis and M.~Salehi, \emph{Digital {Communications}}, 5th~ed.\hskip 1em
  plus 0.5em minus 0.4em\relax McGraw Hill, Nov. 2007.

\bibitem{kschischang_factor_2001}
F.~R. Kschischang and H.-A. Loeliger, ``\BIBforeignlanguage{en}{Factor graphs
  and the sum-product algorithm},'' \emph{\BIBforeignlanguage{en}{{IEEE} Trans.
  Inf. Theory}}, vol.~47, no.~2, pp. 498--519, Feb. 2001.

\bibitem{colavolpe_siso_2011}
G.~Colavolpe, D.~Fertonani, and A.~Piemontese, ``{SISO} detection over linear
  channels with linear complexity in the number of interferers,'' \emph{{IEEE}
  J. Sel. Topics Signal Process.}, vol.~5, no.~8, pp. 1475--1485, 2011.

\bibitem{schmid_low-complexity_2022}
L.~Schmid and L.~Schmalen, ``Low-complexity near-optimum symbol detection based
  on neural enhancement of factor graphs,'' \emph{{IEEE} Trans. Commun.}, pp.
  7562--7575, Nov. 2022.

\bibitem{dempster_maximum_1977}
A.~P. Dempster, N.~M. Laird, and D.~B. Rubin, ``\BIBforeignlanguage{en}{Maximum
  {likelihood} from {incomplete} {data} {via} the \textit{{EM}} {algorithm}},''
  \emph{\BIBforeignlanguage{en}{J. Roy. Stat. Soc.: Ser. B}}, vol.~39, no.~1,
  pp. 1--38, Sep. 1977.

\bibitem{schmid2024blind}
L.~Schmid, T.~Raviv, N.~Shlezinger, and L.~Schmalen, ``Blind channel estimation
  and joint symbol detection with data-driven factor graphs,'' \emph{arXiv
  preprint arXiv:2401.12627}, Jan. 2024.

\bibitem{shlezinger2022discriminative}
N.~Shlezinger and T.~Routtenberg, ``Discriminative and generative learning for
  linear estimation of random signals [lecture notes],'' \emph{{IEEE} Signal
  Process. Mag.}, vol.~40, no.~6, pp. 75--82, 2023.

\bibitem{kingma_adam_2015}
D.~P. Kingma and J.~Ba, ``Adam: {A} method for stochastic optimization,'' in
  \emph{Proc. {Int}. {Conf}. {Learn}. {Represent}. (ICLR)}, San Diego, CA, USA,
  May 2015.

\bibitem{alvarado_achievable_2018}
A.~Alvarado, T.~Fehenberger, B.~Chen, and F.~M.~J. Willems, ``Achievable
  information rates for fiber optics: {Applications} and computations,''
  \emph{J. Lightw. Technol.}, vol.~36, no.~2, pp. 424--439, Jan. 2018.

\bibitem{murphy_loopy_1999}
K.~Murphy, Y.~Weiss, and M.~I. Jordan, ``Loopy belief propagation for
  approximate inference: {An} empirical study,'' in \emph{{Proc}. {Int.}
  {Conf.} {Uncertainty} {Artif.} {Intell.} (UAI)}, Stockholm, Sweden, 1999, pp.
  467--475.

\bibitem{nachmani_deep_2018}
E.~Nachmani, E.~Marciano, L.~Lugosch, W.~J. Gross, D.~Burshtein, and Y.~Beery,
  ``\BIBforeignlanguage{en}{Deep learning methods for improved decoding of
  linear codes},'' \emph{\BIBforeignlanguage{en}{{IEEE} J. Sel. Topics Signal
  Process.}}, vol.~12, no.~1, pp. 119--131, Feb. 2018.

\end{thebibliography}
\end{document}